\documentclass[twocolumn,showpacs,preprintnumbers,amsmath,amssymb,superscriptaddress]{revtex4-1}

\usepackage{graphicx}
\usepackage{dcolumn}
\usepackage{bm}
\usepackage{hyperref}

\begin{document}


\title{The radiative width of the Hoyle state from  $\gamma$-ray spectroscopy}
   \author{T.~Kib\'edi}%
   \email{Tibor.Kibedi@anu.edu.au}
   \affiliation{Department of Nuclear Physics, Research School of Physics, The Australian National University,
      Canberra, ACT, Australia}

\author{B.~Alshahrani}
   \altaffiliation[Present address ]{Department of Physics, Faculty of Science,
   King Khalid University, Abha,
   Kingdom of Saudi Arabia}
  \affiliation{Department of Nuclear Physics, Research School of Physics, The Australian National University,
  Canberra, ACT, Australia}
   \affiliation{Department of Physics, Faculty of Science, King Khalid University, Abha,
   Kingdom of Saudi Arabia}

\author{A.E.~Stuchbery}
  \affiliation{Department of Nuclear Physics, Research School of Physics, The Australian National University,
  Canberra, ACT, Australia}

\author{A.C.~Larsen}
  \affiliation{Department of Physics, University of Oslo, Oslo, Norway}

\author{A.~G\"orgen}
  \affiliation{Department of Physics, University of Oslo, Oslo, Norway}

\author{S.~Siem}
  \affiliation{Department of Physics, University of Oslo, Oslo, Norway}

\author{M.~Guttormsen}
  \affiliation{Department of Physics, University of Oslo, Oslo, Norway}

\author{F.~Giacoppo}
  \altaffiliation[Present address ]{GSI Helmholtzzentrum f\"ur Schwerionenforschung, Darmstadt, Germany, and
   Helmholtz-Institut Mainz, Mainz, Germany}
  \affiliation{Department of Physics, University of Oslo, Oslo, Norway}

\author{A.I.~Morales}
  \altaffiliation[Present address ]{IFIC, CSIC-Universitat de Val\'encia, Val\'encia, Spain}
  \affiliation{Dipartimento di Fisica dell'Universit\'a degli Studi di Milano and INFN-Milano, Milano, Italy}

\author{E.~Sahin}
  \affiliation{Department of Physics, University of Oslo, Oslo, Norway}

\author{G.M.~Tveten}
  \affiliation{Department of Physics, University of Oslo, Oslo, Norway}

\author{F.L.~Bello~Garrote}
  \affiliation{Department of Physics, University of Oslo, Oslo, Norway}

\author{L.~Crespo~Campo}
  \affiliation{Department of Physics, University of Oslo, Oslo, Norway}

\author{T.K.~Eriksen}
  \affiliation{Department of Physics, University of Oslo, Oslo, Norway}

\author{M.~Klintefjord}
  \affiliation{Department of Physics, University of Oslo, Oslo, Norway}

\author{S.~Maharramova}
  \affiliation{Department of Physics, University of Oslo, Oslo, Norway}

\author{H.-T.~Nyhus}
  \affiliation{Department of Physics, University of Oslo, Oslo, Norway}

\author{ T.G.~Tornyi}
  \altaffiliation[Present address ]{Institute of Nuclear Research, MTA ATOMKI, Debrecen, Hungary}
  \affiliation{Department of Physics, University of Oslo, Oslo, Norway}
  \affiliation{Institute of Nuclear Research, MTA ATOMKI, Debrecen, Hungary}

\author{T.~Renstr{\o}m}
  \affiliation{Department of Physics, University of Oslo, Oslo, Norway}

\author{W.~Paulsen}
  \affiliation{Department of Physics, University of Oslo, Oslo, Norway}

\date{\today}

\begin{abstract}
 The cascading 3.21 MeV and 4.44 MeV electric quadrupole transitions have been observed
 from the Hoyle state at 7.65 MeV excitation energy in $^{12}$C, excited by the
 $^{12}$C(p,p$^{\prime}$) reaction at 10.7 MeV proton energy.
 From the proton-$\gamma$-$\gamma$ triple coincidence data, a value of
 ${\Gamma_{\rm rad}}/{\Gamma}=6.2(6) \times 10^{-4}$ was obtained for the radiative branching ratio.
 Using our results, together with ${\Gamma_{\pi}^{E0}}/{\Gamma}$ from Eriksen et al.
 \cite{2020Eriksen_PRC} and
 the  currently adopted $\Gamma_{\pi}(E0)$ values, the  radiative width of the Hoyle state is
 determined as $\Gamma_{\rm rad}=5.1(6) \times 10^{-3}$ eV.
 This value is about 34\% higher than the currently adopted value and will impact on models of stellar
 evolution and nucleosynthesis.
\end{abstract}

\maketitle

\
 The \emph{triple-alpha} reaction, which produces stable $^{12}$C in the universe, is
 a fundamental processes of helium burning stars.
 The entry state of the triple-alpha process, the second excited state in $^{12}$C,
 is a $0^{+}$ state at 7.65 MeV.
 It has attracted  significant attention  \cite{2014Fr14,2020Sm01,2016Fu07}
 since it was first proposed in 1953 by Fred Hoyle \cite{1953Ho81}.
 The existence of the state was confirmed in the same year from the analysis of the
 $\alpha$-spectrum from the $^{14}$N(d,$\alpha$)$^{12}$C reaction  \cite{1953Du23}.
 The Hoyle state is $\alpha$ unbound and the dominant decay process ($>99.94$\%) is
 through the emission of an alpha particle, leading to the very short lived isotope, $^8$Be,
 which then disintegrates into two alpha particles.
 Stable carbon will only be produced either if the Hoyle state decays directly to the ground
 state via  an electric monopole (E0) transition or  by a cascade
 of two electric quadrupole (E2) transitions.

 Due to its  unusual structure, the Hoyle  state has attracted
 continuous attention; see the recent review of
 Freer and Fynbo \cite{2014Fr14} and other recent works
 \cite{2013Zi03,2015Fu09,2020Sm01}.
 The discussion includes nuclear clustering, a spacial arrangement of the three $\alpha$ particle
 clusters of which the state is believed to be composed, and discussion on a new form of nuclear matter,
 in analogy with the Bose-Einstein condensates.
 The characterization of the $2^+$ and $4^+$ states on top of the 7.65 MeV $0^+$ state,
 forming the \emph{Hoyle band} \cite{2018Garg_JPC}, together with much improved \emph{ab initio}
 calculations \cite{2018Launey_AIP} are important steps forward.

 The production rate of stable carbon in the universe is cardinal for many aspects
 of nucleosynthesis.
 The reaction rate is closely related to the decay properties of the Hoyle state.
 The triple-alpha reaction rate can be expressed as:
  $r_{3\alpha} = \Gamma_{\rm rad} \exp(-Q_{3\alpha}/kT)$ \cite{1988RolfsRodney}.
 Here $\Gamma_{\rm rad}$ is the total electromagnetic (radiative) decay width,
 $Q_{3\alpha}$ is the energy release in the three $\alpha$ breakup of the Hoyle state, and
 \emph{T} is the stellar temperature.
 $\Gamma_{\rm rad}$ has contributions from the 3.21-MeV E2 and the 7.65-MeV E0  transitions.
 The contributions of electron conversion are negligible, so including photon ($\gamma$)
 and pair conversion ($\pi$),
 $\Gamma_{\rm rad} = \Gamma_{\gamma}^{E2} + \Gamma_{\pi}^{E2} +\Gamma_{\pi}^{E0}$.
 Based on current knowledge, 98.4\% of the electromagnetic decay width
 is from the E2 photon emission and 1.5\% is from the E0 pair decay \cite{2009KiZZ}.
 The $\Gamma_{\pi}^{E2}$ contribution is less than 0.1\%.

 The value of $\Gamma_{\rm rad}$ cannot be directly measured.
 It is usually evaluated as a product of three independently measured quantities:
 \begin{equation}\label{Eqn:Gamma_rad}
  \Gamma_{\rm rad} = \left[ \frac{\Gamma_{\rm rad}}{\Gamma} \right] \times
  \left[ \frac{\Gamma}{\Gamma_{\pi}(E0)} \right] \times
  \left[ \Gamma_{\pi}(E0) \right] \, ,
 \end{equation}
 where $\Gamma$ is the total decay width of the Hoyle state, which includes the
 $\alpha$, as well as the E2 and E0 electromagnetic decays.

 The only absolute quantity in Eq.~(\ref{Eqn:Gamma_rad}) is $\Gamma_{\pi}(E0)$,  which
 has been measured 8 times
 \cite{1956Fr27, 1964Cr01,1965Gu04,1967Cr01,1968St20,1970St10,2005Cr03,2010Ch17}.
 The two most recent measurements \cite{2005Cr03,2010Ch17} are the most precise;
 however they disagree by more than 5$\sigma$.
 Following the recommendation of Freer and Fynbo \cite{2014Fr14}, we have adopted a
 value of 62.3(20) $\mu$eV from the latter study.

 The least precisely known quantity is ${\Gamma_{\pi}(E0)}/{\Gamma}$.
 Combining all previous measurements \cite{1960Al04,1960Aj04,1972Ob01,1977Al31,1977Ro05}, a value of
 ${\Gamma_{\pi}(E0)}/{\Gamma}=6.7(6) \times 10^{-6}$ was adopted \cite{2014Fr14}.
 This value has been further improved by a new pair conversion measurement at the ANU
 \cite{2020Eriksen_PRC} and a ${\Gamma_{\pi}(E0)}/{\Gamma}$ ratio of $7.6(4) \times 10^{-6}$
 was recommended.

 The third term, ${\Gamma_{\rm rad}}/{\Gamma}$, has been measured 8 times between 1961 and 1976
 \cite{1961Al23,1963Se23,1964Ha23,1974Ch03,1975Da08,1975Ma34,1976Ma46,1976Ob03}.
 By excluding the value of  2.8(3)$\times 10^{-4}$ by Seeger and Kavanagh \cite{1963Se23},
  the weighted mean value is 4.13(11)$\times 10^{-4}$.
 In Ref.~\cite{2014Fr14}, a slightly higher value of
 4.19(11)$\times 10^{-4}$ was recommended.
 ${\Gamma_{\rm rad}}/{\Gamma}$ is claimed to be the most precise term in
 Eq.~(\ref{Eqn:Gamma_rad}).

 In the present paper we  report a new measurement of ${\Gamma_{\gamma}^{E2}}/{\Gamma}$,
 which was deduced from the rate of proton-$\gamma$-$\gamma$ triple coincidences, $N^{7.65}_{020}$,
 corresponding to the de-excitation of the Hoyle state through the emission of the 3.21 and 4.44 MeV
  $\gamma \gamma$-cascade, to the rate of singles proton events,
 $N^{7.65}_{\rm singles}$, exciting the Hoyle state:
  \begin{equation}
  \frac{\Gamma_{\gamma}^{E2}}{\Gamma}=\frac{N^{7.65}_{020}}
  {N^{7.65}_{\rm singles} \times \epsilon_{3.21} \times \epsilon_{4.44} \times W^{7.65}_{020}}  \, ,
  \label{eqn:Gg_G_direct}
  \end{equation}
 where
 $\epsilon_{3.21}$ and $\epsilon_{4.44}$ are the photon detection efficiencies, and
 $W^{7.65}_{020}$ is the angular correlation correction for a 0-2-0 cascade.
 Our approach is similar to that of Obst and Braithwaite \cite{1976Ob03},
  but with much improved experimental apparatus and analysis techniques.

\begin{figure}[]
\vspace*{0pt}
\includegraphics[width=0.42\textwidth]{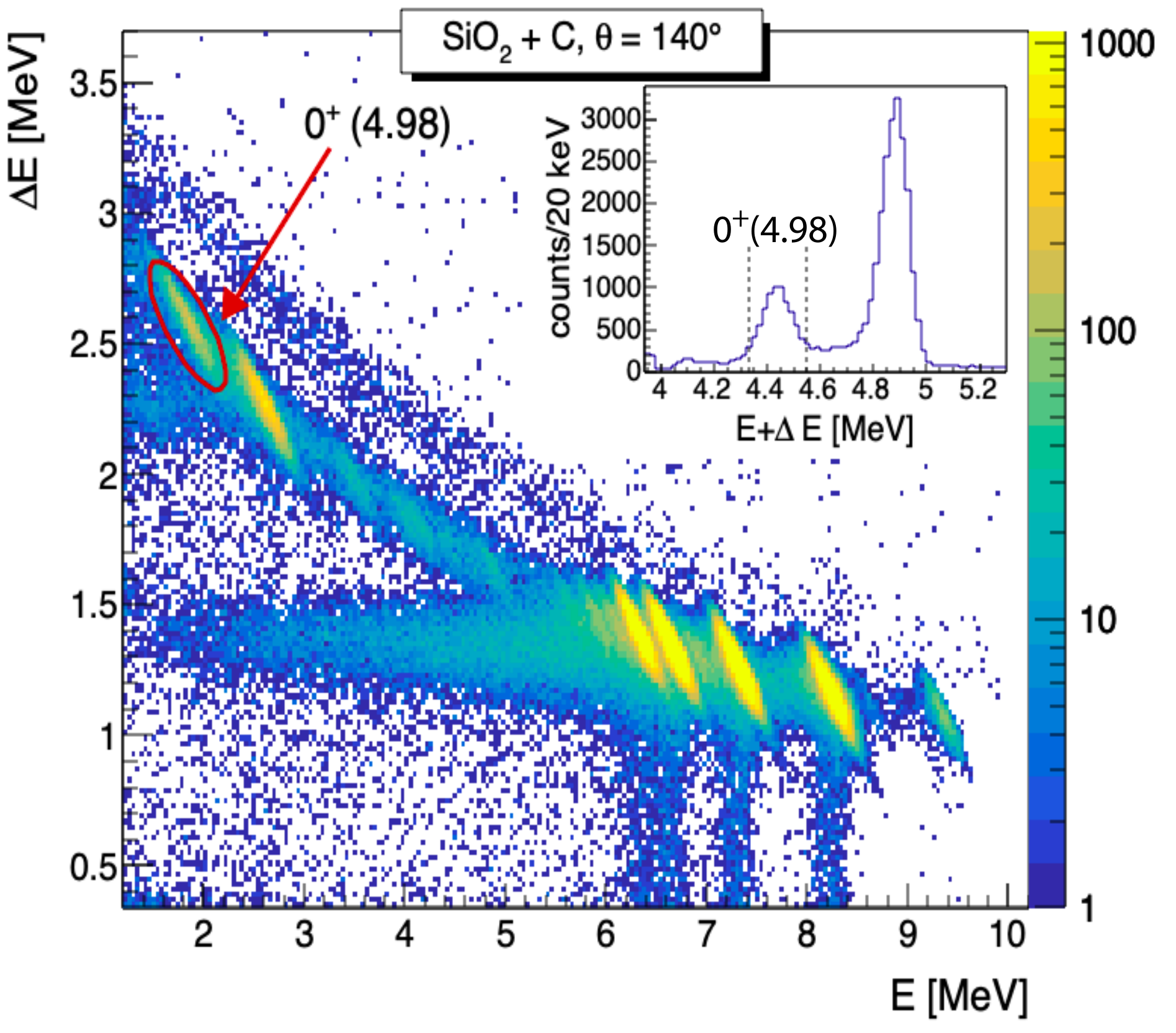} 
\vspace*{-10pt}
\caption{\label{fig:28Si_EDE}
  Singles proton events recorded in the SiRi $E$ (horizontal axis) vs. $\Delta E$ (vertical axis)
  telescopes using the SiO$_{2}$  plus carbon target.
  Events corresponding to the excitation of the 4.98 MeV $0^{+}$ state are
  indicated by the ellipse.
  The insert shows $\Delta E+E$ total energy spectrum around the proton group of the 4.98 MeV
  $0^{+}$ state, together with the energy gate (dashed lines). }
\end{figure}

 The experiments were carried out at the Cyclotron Laboratory of the University of Oslo.
 The Hoyle state was populated in inelastic scattering of 10.7 MeV protons on a 180 $\mu$g/cm$^2$
 natural carbon target.
 This energy was slightly higher than the notional optimum energy of 10.5 MeV, where the
 $^{12}$C(p,p$^{\prime}$) reaction has a relatively broad resonance \cite{1971Da36}.
 The higher proton energy was employed to shift the inelastically scattered protons to $\sim$1.5 MeV,
  well above the detecting threshold.
  Proton angular distribution measurements \cite{2020Eriksen_PRC} suggest that the
  ratio of the excitation of the 4.44 MeV and 7.65 MeV states is essentially the same  at 10.5
  and 10.7 MeV.
  In the present experiments a beam intensity of 5 nA was used, keeping the total count rate
  below 3 kHz.
  Additional experiments were carried out using a target consisting of a layer of
  140 $\mu$g/cm$^2$ SiO$_2$ on a 32 $\mu$g/cm$^2$ natural carbon backing.
  The $^{28}$Si(p,p$^{\prime}$) reaction was used to determine the photon detection efficiencies.
   The $0^{+}$ state at 4.98 MeV in $^{28}$Si decays with a 100\% branching ratio to the ground state
  through the emission of a 3.20 MeV  - 1.78 MeV cascade.
  In addition, the 4.50 MeV  - 1.78 MeV  cascade from the 6.28 MeV $3^+$ state was also analyzed.
  The branching ratio of this cascade is $BR^{6.28}_{\gamma}=88.2(4)\%$ \cite{2013Ba53}.

\begin{figure}[]
\includegraphics[width=0.36\textwidth,angle=0]{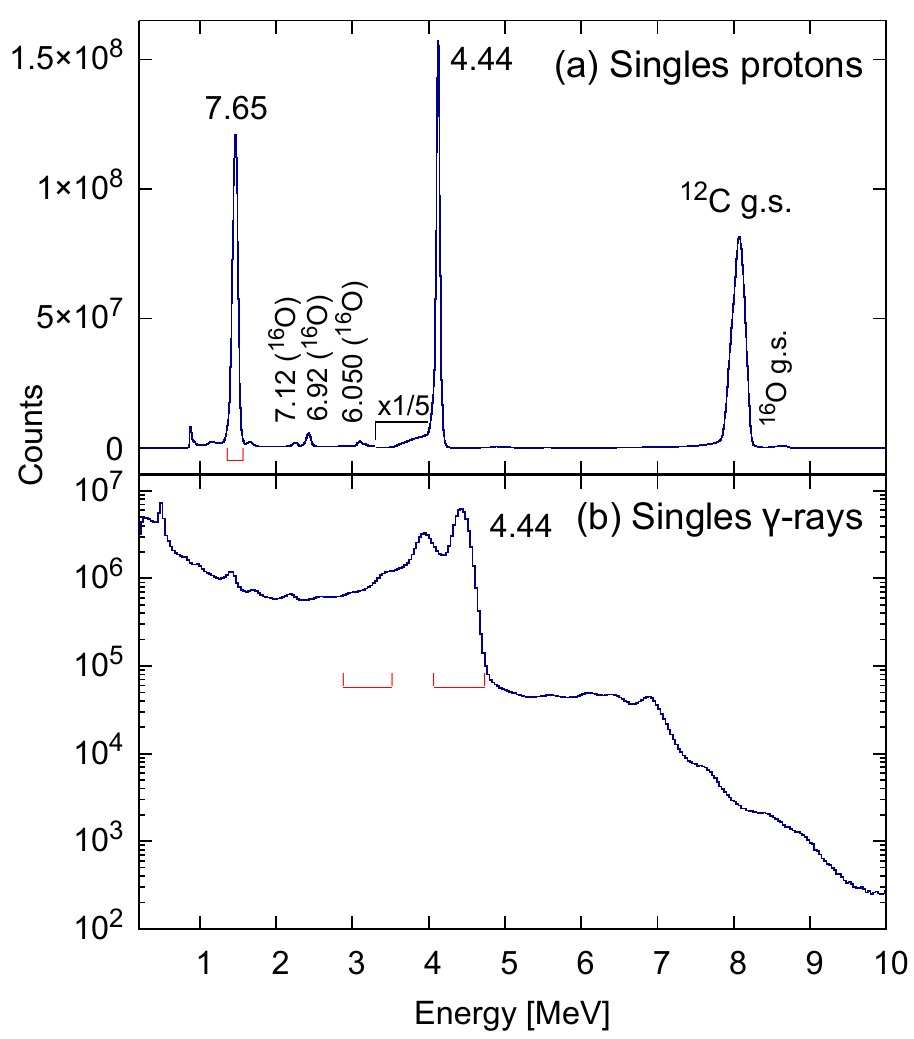} 
\vspace*{-12pt}
\caption{\label{fig:12C_Singles}
 Singles spectra of
  \textit{(a)} protons and \textit{(b)} $\gamma$-rays using the $^{12}$C(p,p$^{\prime}$) reaction.
  The proton ($7.65_{\rm p}$) and $\gamma$-ray energy ($3.21_{\gamma}$ and $4.44_{\gamma}$)
  gates used for the analysis are indicated by red lines.
  }
\end{figure}

  Proton-$\gamma$-$\gamma$ coincidences were measured with the SiRi particle telescope
  \cite{2011Guttormsen} and the  CACTUS $\gamma$-ray detector array \cite{1990Guttormsen}.
  The 64  $\Delta E - E$ telescopes of SiRi were placed in the backward direction covering
  angles between $126^{\circ}$ and $140^{\circ}$ relative to the beam direction.
  The solid angle of the particle detection was  around 6\% of 4$\pi$.
  The front ($\Delta E$) and back ($E$) particle detectors have thicknesses of $130 \, \mu$m
  and $1550 \, \mu$m, respectively.
  $\gamma$-rays were recorded with the CACTUS array consisting of 26 collimated 5" $\times$ 5" NaI(Tl)
  detectors, placed at 22 cm from the target.
  Each detector had a 10 cm lead collimator to ensure illumination of the center
  of the detector.
  The total photon efficiency of the array is $\approx$14.2\% of 4$\pi$ at 1.33 MeV energy.

  Signals in the $\Delta E$ detectors were used as triggers and to start the
  time-to-digital-converter (TDC).
  The stop signal was generated when any NaI(Tl) detector fired.
  In this way prompt proton-$\gamma$-$\gamma$ coincidences
  could be sorted from the event-by-event data.
  Fig.~\ref{fig:28Si_EDE} shows the energy deposition in the $E$ vs. $\Delta E$ detectors recorded with
  the SiO$_{2}$ plus carbon target.
  The fraction of the particle energy deposited in the front detector depends on $Z$, $A$ and the
  particle energy.
  This relation, visible in Fig.~\ref{fig:28Si_EDE} as a ``banana" shaped region, can be used to
  identify the detected particles, and also to filter events of incomplete energy deposition
  (horizontal and vertical bands), as well as other beam related background events.
  The $\Delta E-E$ spectrum can be used to select the population of specific states.
  Protons exciting the Hoyle state fully stop in the $\Delta E$ detector.
  In this case the $\Delta E-E$ telescope was operated in anti-coincidence to reject high energy
  particle  events depositing only partial energy in the $\Delta E$ detector.

  \begin{figure}[]
\includegraphics[width=0.48\textwidth,angle=0]{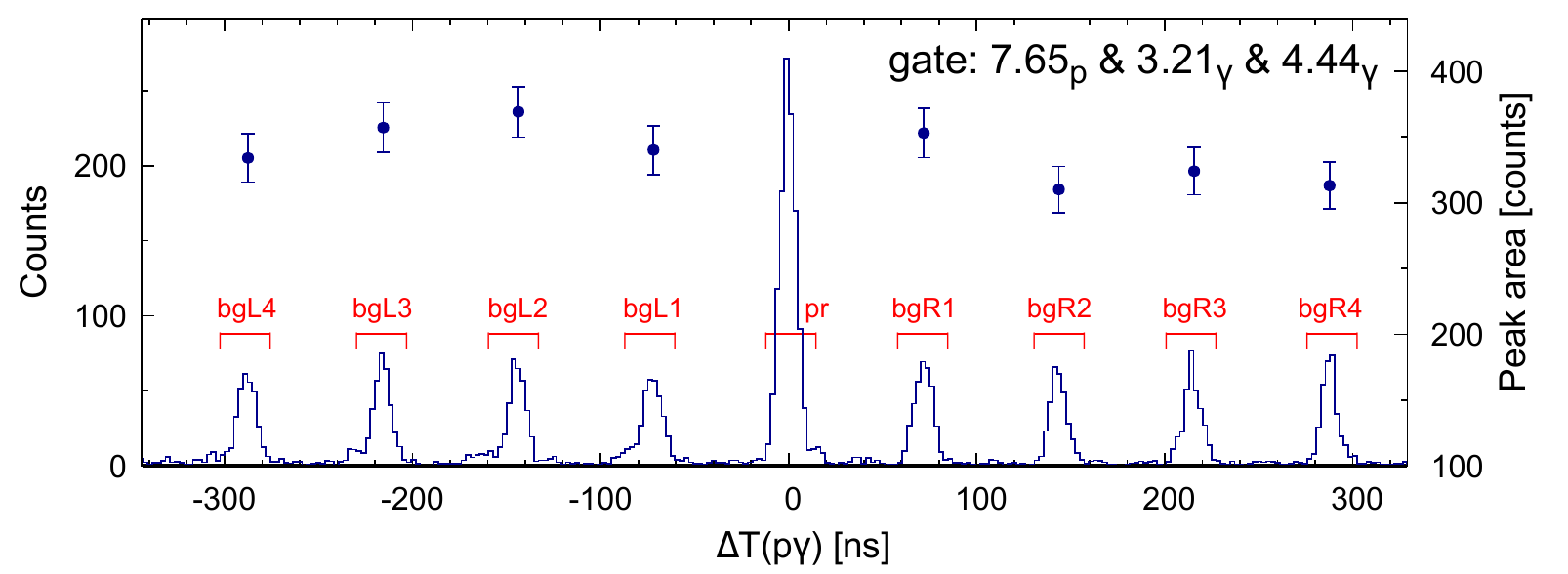} 
\vspace*{-18pt}
\caption{\label{fig:12C_7.65_TDC}
 Time differences between protons exciting the Hoyle state and 3.21- and 4.44-MeV $\gamma$-rays.
 The prompt (\emph{pr}) and four background gates on each side (\emph{bgLx}, \emph{bgRx})
 are marked in red.
 The average counts in the background peaks is 318(18).
  }
\end{figure}

  Fig.~\ref{fig:12C_Singles} shows the spectra of singles proton and $\gamma$-ray events from the
  $^{12}$C(p,p$^{\prime}$) reaction collected over a period of 12 days.
  The peak at 1.5 MeV proton energy, labelled as ``7.65", represents the excitation of the Hoyle state.
  It  contains  $N_{\rm singles}^{7.65} = 2.78(6)\times 10^8$ events, however only 1 out of
  $\sim$2500 proton excitations is expected to result in electromagnetic transitions leading to
  the ground  state of $^{12}$C.
  In comparison, the number of protons exciting the 4.44 MeV $2^+$ state is about 4.7 times higher, and
  this state always decays to the ground state with an E2 $\gamma$-ray transition.
  The singles $\gamma$-ray spectrum, shown in panel (b) of Fig.~\ref{fig:12C_Singles}, is
  dominated by the 4.44 MeV photon events.
  Beside the full energy peak, there is a broad
  distribution of events of single (at $\sim$3.9 MeV) and double (at $\sim$3.4 MeV) escapes, as well
  as Compton scattering.
  The 3.21 MeV transition is expected to be about 10000 times weaker
  and it partially overlaps with  second escape peak of the 4.44 MeV line.
  In this energy region the photon energy resolution was around 0.19 MeV.
  Excitation of the 4.44 MeV state will only produce a single photon event.
  However, we estimated that the probability of two 4.44 MeV $\gamma$-rays produced by two unrelated reactions
  and observed in  prompt coincidence is $7 \times 10^{-5}$ per second, which is about three times
  lower than the true coincidence rate and can be considered as high.

  Fig.~\ref{fig:12C_7.65_TDC} shows the time differences between protons exciting the Hoyle state and
  a pair of 3.21 and 4.44 MeV $\gamma$-rays.
  The main peak at $\Delta T(p\gamma)$= 0 ns (\emph{``pr"}) corresponds to $\gamma$-rays in
  prompt coincidence with protons.
  The secondary peaks (\emph{``bgLx"} and \emph{``bgRx"}) occurring every 72 ns
  are from accidental coincidences where one of the two gamma-rays was produced in another beam burst.
  The 4 background gates either side of the prompt and equal width to the prompt peak were averaged over.

\begin{figure}[]
\includegraphics[width=0.40\textwidth,angle=0]{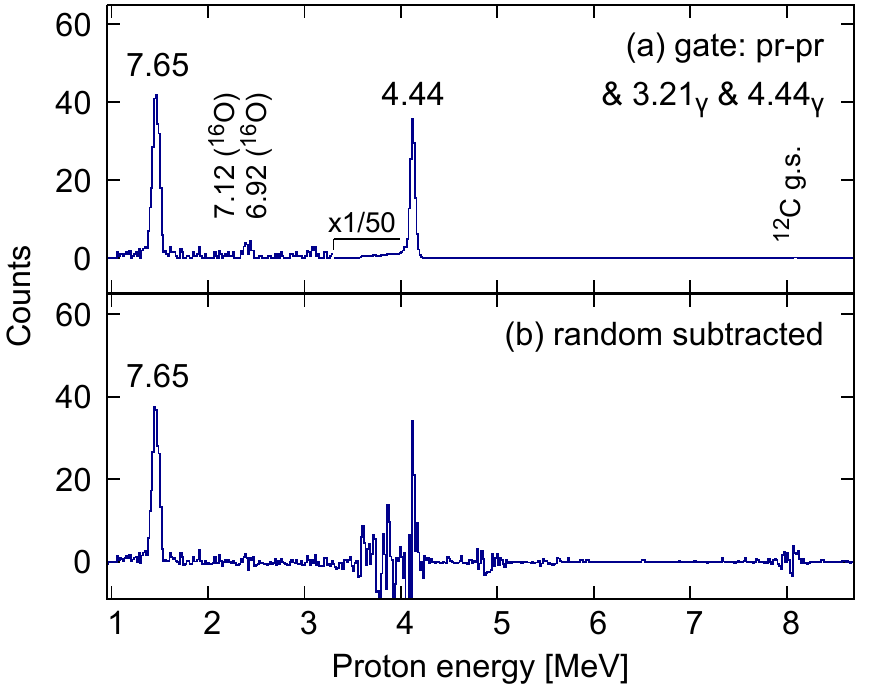}
\vspace*{-8pt}
\caption{\label{fig:12C_7.65_DEE}
 Protons in prompt coincidence with  3.21 and 4.44 MeV $\gamma$-rays cascade.
 Panel \emph{(a)}: both $\gamma$-rays observed in the in prompt (\emph{``pr-pr"}) TDC window;
 \emph{(b)}:  random events from the \emph{``pr-bgLx"} and \emph{``pr-bgRx"} TDC
 gates (Fig.~\ref{fig:12C_7.65_TDC}) are subtracted.
 }
\end{figure}

  Panel (a) of Fig.~\ref{fig:12C_7.65_DEE} shows protons (``7.65") in prompt coincidence with a 3.21
  and a 4.44 MeV $\gamma$-rays without subtraction of accidental coincidences.
  In the same spectrum $N_{p}(2_{1}^{+})$, the number of protons exciting the
  $2^{+}_{1}$ state (``4.44"), is due to
  accidental coincidences and is nearly 50 times higher.
  Using TDC gates of \emph{``pr-pr"}, \emph{``pr-bgLx"}, \emph{``pr-bgRx"} and \emph{``bgLx-bgRx"}
  the numbers of $N_{p}(2^{+}_{1})$ events in the corresponding
  proton spectra are 8251(91), 7697(88), 7914(89) and 54(9), respectively.
  Protons exciting the $2^{+}_{1}$ state will only produce single photon events, therefore the
  $N_{p}(2^{+}_{1})$ rates can be used to  remove the random events.
  Using the above $N_{p}(2^{+}_{1})$ rates
  the  scaling factor was obtained as $8251(91) / [[7697(88)+7914(89)- 54(9)]/2] = 1.061(12)$.
  The $N_{p}(0^+)$ rates in the same TDC gates were
  249(16), 158(13), 197(14) and 66(8), respectively.
  This gives $N^{7.65}_{020} = 212(22)$  counts.
  The final proton spectrum in triple
  coincidence with the 3.21 and 4.44 MeV $\gamma$-rays is shown in panel (b) of Fig.~\ref{fig:12C_7.65_DEE}.

\begin{figure}[]
\includegraphics[width=0.50\textwidth,angle=0]{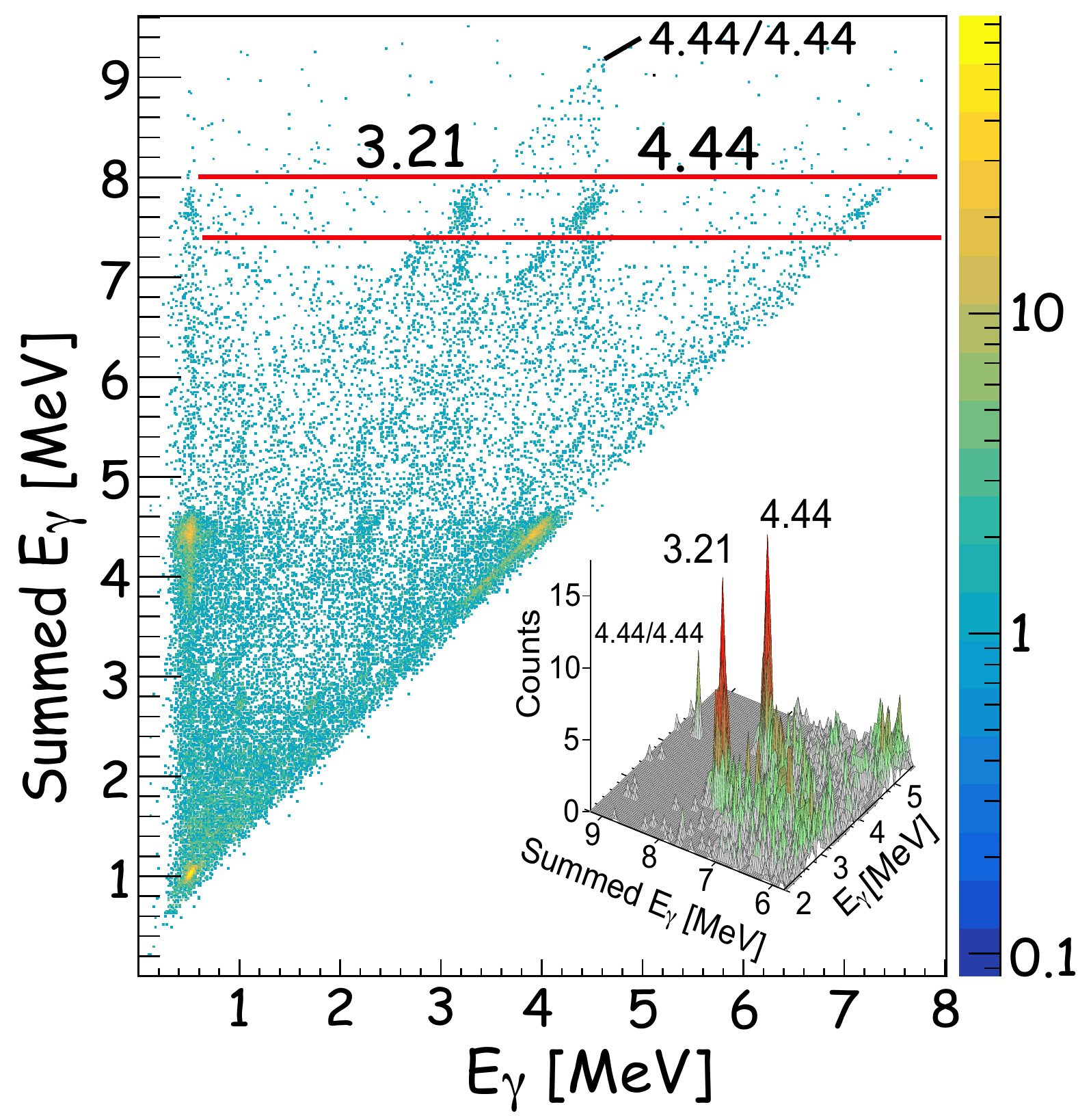}
\vspace*{-18pt}
\caption{\label{fig:12C_7.65_G_sG_ROOT}
 $\gamma$-ray energy vs. summed $\gamma$-ray energy matrix constructed from $\gamma$-$\gamma$ coincidence events gated
 by protons exciting the Hoyle state.
 Random events have been removed.
 The gate representing the 3.21 plus 4.44 MeV summed energy ($7.65_{\rm sum}$) is indicated with red horizontal lines.
 The insert shows the region around the 3.21 and 4.44 MeV transitions in 3D.
 Data have been compressed by factor 4.
 The location of the random coincidences of the 4.44 MeV $\gamma$-ray with itself is also marked.
  }
\end{figure}

 Fig.~\ref{fig:12C_7.65_G_sG_ROOT} shows the $\gamma$-$\gamma$ coincidence events gated by protons
 exciting the Hoyle state, where the horizontal axis is the $\gamma$-ray energy and the vertical axis
 is the summed energy of the two gamma-rays in coincidence.
 The number of random events has been evaluated using the accidental coincidences of the 4.44 MeV
 gamma-ray with itself, indicated as ``4.44/4.44".
 The number of such events in the various TDC gates
  were 131(12), 157(13), 134(12), 63(8),
 which gives a subtraction factor of 1.15(11), a value consistent with the one obtained from the
  proton spectra.
  To deduce the final $\gamma \gamma$ coincidence spectra, the scaling factor of 1.061(12) was adopted.
  Fig.~\ref{fig:12C_7.65_G_sG_ROOT} also shows the final matrix of $\gamma \gamma$
  coincidence events.
  A small residue of the 4.44-4.44 random coincidences is visible, but the number of related
  events under the peaks of interest is negligible.

\begin{figure}[]
\includegraphics[width=0.42\textwidth,angle=0]{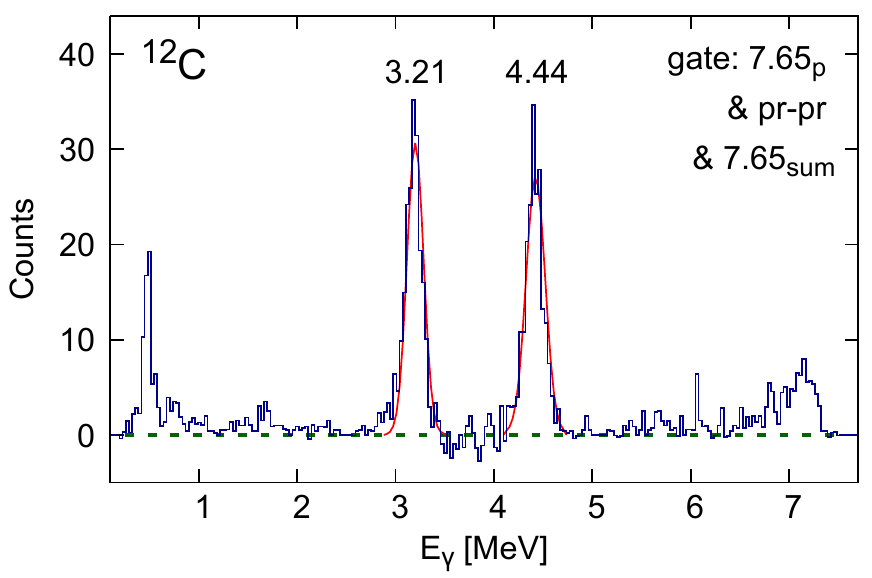}
\vspace*{-10pt}
\caption{\label{fig:12C_7.65_pGG}
 Random subtracted $\gamma$-rays from the Hoyle state.
 The fit to the spectrum including the 3.21 and 4.44 MeV transitions is shown in red.
  }
\end{figure}

 The final $\gamma$-ray spectrum of the 3.21-4.44 MeV cascade is shown in
 Fig.~\ref{fig:12C_7.65_pGG}.
 The areas of the 3.21 and 4.44 MeV photon peaks, 208(21) and 213(21) counts, were obtained
  by fitting Gaussian functions to these data.

  Using the scaling factor of 1.061(12), the true triple coincidence
  events in the prompt $p \gamma$ peak in Fig.~\ref{fig:12C_7.65_TDC} was evaluated as
  $N_{020}^{7.65}=237(23)$.
  The adopted value of the $N_{020}^{7.65}=217(21)$ was obtained as the weighted mean of the three values
  deduced from the different projections.

  The absolute photon detection efficiency, $\epsilon$, was evaluated
  using the Penelope code \cite{2008Savat}.
  The same simulations were used to evaluate the correction factors,
  $W_{020}$ and $W_{320}$, for the $\gamma$-ray angular correlation, including geometrical
  attenuation coefficients \cite{1953Rose}, listed in Table~\ref{tab:qunatities}.
  To confirm the accuracy of the simulations, the proton gated spectrum of
  the 1.78 and 4.50 MeV $\gamma$-rays from the 6.28 MeV $3^{+}$ state in $^{28}$Si was used.
  The ratio of the peak areas of the 1.78 MeV and 4.50 MeV transitions is 1.58(3), which
  after applying the 1.0170(15) correction for the angular correlation,
  is very close to the value of 1.63(4) from the simulations.

  By evaluating Eq.~\ref{eqn:Gg_G_direct} with values from Table~\ref{tab:qunatities} and considering
  all 325 NaI detector combinations, we obtained $\Gamma_{\gamma}^{E2}/{\Gamma}=6.1(6) \times 10^{-4}$.

  To reduce dependence on the Monte Carlo evaluation of the absolute efficiencies and perform an
  analysis similar to that of Obst and Braithwaite \cite{1976Ob03}, the
  $\Gamma_{\gamma}^{E2}/{\Gamma}$ ratio was deduced using:
  \begin{equation}
  \frac{\Gamma_{\gamma}^{E2}}{\Gamma} = \frac{N^{7.65}_{020}}{N^{4.98}_{020}} \times
        \frac{N^{4.98}_{\rm singles}}{N^{7.65}_{\rm singles}} \times
        \frac{\epsilon^{1.78}_{\gamma}}{\epsilon^{4.44}_{\gamma}} \times
        \frac{\epsilon^{3.20}_{\gamma}}{\epsilon^{3.21}_{\gamma}} \times
        \frac{W^{4.98}_{020}}{W^{7.65}_{020}}\, .
   \label{eqn:Gradg_G}
  \end{equation}
  The symbols are as given for Eq.~(\ref{Eqn:Gamma_rad}).
  An alternative equation can be obtained using the 6.28 MeV $3^+$ state in $^{28}$Si.
  Using the singles proton and $p \gamma \gamma$ triple coincidence rates of the 4.98 MeV and
  6.28 MeV states, the ratio of the proton to photon efficiencies could be determined.
  Combining the results from Eq.~\ref{eqn:Gradg_G} and using numerical values from
  Table~\ref{tab:qunatities}, we again obtain $\Gamma_{\gamma}^{E2}/{\Gamma}$=6.1(6)$\times 10^{-4}$.

  Using the theoretical total conversion coefficient,
  $\alpha_{tot}(E2, 3.21 \, {\rm MeV})=8.77(13) \times 10^{-4}$ \cite{2008Ki07} and
  the recommended value of ${\Gamma_{\pi}(E0)}/{\Gamma}$ \cite{2020Eriksen_PRC},  we obtain
 $\Gamma_{\rm rad}/\Gamma = 6.2(6) \times 10^{-4}$.
  This value is more than 3$\sigma$ away from the currently recommended $\Gamma_{\rm rad}/\Gamma$
  value \cite{2014Fr14}.
  Most of the previous measurements \cite{1974Ch03,1975Da08,1975Ma34,1976Ma46} were based on
  counting the number of $^{12}$C atoms surviving after the Hoyle state was formed  in
  various nuclear reactions.
  To achieve high statistics, the particle detection was carried out
   without magnetic selection and often with reported count rates  above 10 kHz.
  Under these conditions the elimination of accidental coincidences is very challenging.

  The investigation by Obst and Braithwaite \cite{1976Ob03} deduced the
  $\Gamma_{\gamma}^{E2}/\Gamma$ ratio using a similar procedure to the present study.
  Their final result, which was obtained using Eq.~(14) of their paper, contains five
  ratios ($A$ to $E$).
  Despite some differences between their experiment and ours, various combinations of
  these ratios should agree within a few percent.
  The largest difference occurs for
  $B \times D = (N^{6.28}_{320} \times N^{4.98}_{\rm singles}) / (N^{6.28}_{\rm singles}
  \times N^{4.98}_{020})$,
  Ref.~\cite{1976Ob03} reports 0.409(15) whereas our value is 0.80(4).
  Thus most of the difference between Obst and Braithwaite \cite{1976Ob03} and our work stems
  from the $N^{4.98}_{020} / N^{4.98}_{\rm singles}$ ratio in the $^{28}$Si calibration data.
   Our results  were independently checked in Canberra and Oslo
   using different analysis software.

 \begin{table}[]
 \caption{\label{tab:qunatities} Quantities used to evaluate
  $\Gamma_{\gamma}^{E2}/{\Gamma}$ ratio.}
 \begin{ruledtabular}

 \begin{tabular}{lccc}
       &
 $0^+$(7.65 ) &
 $0^+$(4.98) &
 $3^+$(6.28)  \\ \hline
 $N_{\rm 020 \, or \, 320}$  &
 217(21)                  & 
 2233(68)                 &
 6295(106)               \\
 $N_{\rm singles}$            &
 $2.78(6) \times 10^{8}$  &
 $1.08(2) \times 10^{6}$  &
 $3.82(8) \times 10^{6}$  \\
 $\gamma$-ray  &
 $\epsilon_{\rm 3.21}$=0.221(3) &
 $\epsilon_{\rm 3.20}$=0.222(3) &
 $\epsilon_{\rm 4.50}$=0.186(3)\\
 \hspace*{2pt} efficiency [\%] &
 $\epsilon_{\rm 4.44}$=0.187(3) &
 \multicolumn{2}{c}{$\epsilon_{\rm 1.78}$=0.304(3)}\\
 $W_{\rm 020 \, or \, 320}$              &
 0.9582(15)                 &
 0.9623(15)                 &
 1.0170(15)                  \\
 \end{tabular}
 \end{ruledtabular}
 \end{table}

   Moreover, the data of Obst and Braithwaite for $B \times D$ are not self consistent.
   Using the photon efficiencies, the correction factors for the $\gamma \gamma$ angular correlations
   and the $\gamma$-ray branching ratio from the $3^{+}$, $BR^{6.28}_{\gamma}$ state we have:
  \begin{equation}
    B \times D \times \frac{\epsilon_{3.20}}{\epsilon_{4.50}}\times \frac{W_{020}^{4.98}}{W_{320}^{6.28}}
     = BR_{\gamma}^{6.28} = 0.882(4) \, .
  \label{eqn:BvsD}
  \end{equation}
  The data of \cite{1976Ob03} are in disagreement  with Eq.~(\ref{eqn:BvsD}) by a factor of two;
  the present data  (Table~\ref{tab:qunatities}) agree within 2\%.

  Finally, using the recommended $\Gamma_{\pi}(E0) / \Gamma$ \cite{2020Eriksen_PRC},
  the adopted $\Gamma_{\pi}^{E0}$ and our  $\Gamma_{\rm rad}/\Gamma$ values, the
  radiative width of the Hoyle state  is $ \Gamma_{\rm rad} = 5.1(6) \times 10^{-3} \, \text{eV}$.
  This result suggests a significantly higher radiative width than currently adopted.

  The \emph{triple-alpha} reaction together with $^{12}$C($\alpha , \gamma$) are the two most
  important helium burning nuclear reactions with a significant impact on nucleosynthesis and
  the evolution of massive stars   \cite{2007Tur_Apj,2013West_Apj,2018Fields_Apj}.
  In the core-He burning cycle these reactions compete to determine the relative carbon and oxygen
  abundances before the core-C burning starts.
  The uncertainties due to  production rates grow  at every step.
  This makes the uncertainty of the \emph{triple-alpha} and the $^{12}$C($\alpha , \gamma$)
  reaction rates crucial for the production  of heavy elements.
  Recent calculations \cite{2013West_Apj,2018Fields_Apj} have explored variations within
   the uncertainties of
  the production rates: $\pm 10 \%$ for the triple-alpha and $\pm 25 \%$ for the
  $^{12}$C($\alpha , \gamma$) reactions.
  West, Heger and Austin \cite{2013West_Apj} pointed out that a 25\% increase
  in the triple alpha rate would be consistent with a 33\% larger $^{12}$C($\alpha , \gamma$) rate.
  Here we report a 34\% change in the triple-alpha reaction rate, which is outside
  of the parameter space of the calculations.
  This scenario needs to be explored, as it could change many
  of the model predictions.

  In summary, a new measurement of the $\Gamma_{\rm rad} / \Gamma$ ratio  of the
  Hoyle state has been performed using a much improved experimental setup than used in the last
  study, more than 40 years ago, giving a value that is significantly higher.
  The accurate determination of the triple-alpha rate remains a challenge for low energy nuclear physics.
  The present experiment only focused on one of the three terms defined in Eq.~(\ref{Eqn:Gamma_rad}).
  Confirmation of the new result, using higher resolution photon spectrometers is well warranted.
  Additional experiments of the $\Gamma_{\pi}(E0) / \Gamma$ ratio, as well as
  of the E0 width, $\Gamma_{\pi}(E0)$ are equally important.\\

  \begin{acknowledgments}
  The project was supported by the Australian Research Council Discovery Grants DP140102986,
  DP170101673 and by the Research Council of Norway, Grant 263030.
  TK, BA and AES acknowledge the hospitality of the University Oslo during the experiments.
  ACL gratefully acknowledges funding from the European Research Council through ERC-STG-2014,
  Grant Agreement no. 637686.
\end{acknowledgments}

\end{document}